# A PHYSICALLY EXTENDED EEIO FRAMEWORK FOR MATERIAL EFFICIENCY ASSESSMENT IN UNITED STATES MANUFACTURING SUPPLY CHAINS

PRE-PRINT

Heather Liddell* – Purdue University
Beth Kelley – Purdue University
Liz Wachs – National Renewable Energy Laboratory
Alberta Carpenter – National Renewable Energy Laboratory
Joe Cresko – U.S. Department of Energy

*Corresponding Author: liddellh@purdue.edu

## Abstract

A physical assessment of material flows in an economy (e.g., material flow quantification) can support the development of sustainable decarbonization and circularity strategies by providing the tangible physical context of industrial production quantities and supply chain relationships. However, completing a physical assessment is challenging due to the scarcity of high-quality raw data and poor harmonization across industry classification systems used in data reporting. Here we describe a new physical extension for the U.S. Department of Energy's (DOE's) EEIO for Industrial Decarbonization (EEIO-IDA) model, yielding an expanded EEIO model that is both physically and environmentally extended. In the model framework, the U.S. economy is divided into goods-producing and service-producing subsectors, and mass flows are quantified for each goods-producing subsector using a combination of trade data (e.g., UN Comtrade) and physical production data (e.g., U.S. Geological Survey). Given that primary-source production data are not available for all subsectors, price-imputation and mass-balance assumptions are developed and used to complete the physical flows dataset with high-quality estimations. The resulting dataset, when integrated with the EEIO-IDA tool, enables the quantification of environmental impact intensity metrics on a mass basis (e.g., $CO_2$eq/kg) for each industrial subsector. This work is designed to align with existing DOE frameworks and tools, including the EEIO-IDA tool, the DOE Industrial Decarbonization Roadmap (2022), and Pathways for U.S. Industrial Transformations study (2025).

## Introduction and Motivation

An economic input-output (IO) model tracks the flow of goods through an economy and can be used to analyze the accrual of value in product supply chains, from raw material extraction to intermediates to final products. This is shown conceptually in Figure 1. These models can offer insights into the overall structure of the economy, including where raw materials from extraction industries enter the economy, the extent to which industries transact with each other in domestic or international supply chains, and how these transactions between industries contribute to cumulative value addition in gross output. These features make IO models ideal for the study of circularity.[1-5]

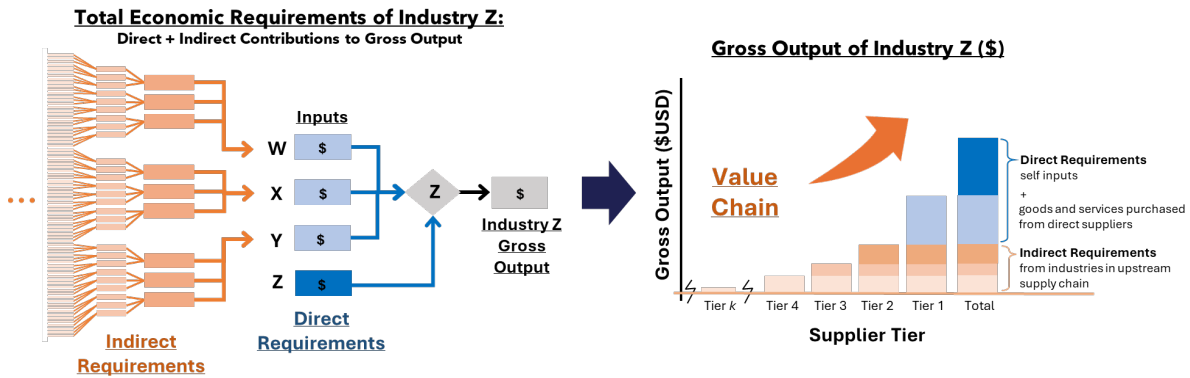

*Figure 1. Schematic representation of the economic IO framework*

An environmentally extended input-output (EEIO) model leverages the economic IO framework (and the comprehensive, high-quality scope of economic trade data) to quantify environmental impacts in supply chains. The economy-level view afforded by an EEIO model allows for the quantification of emissions and other environmental flows for a cross-section of goods and services representing (in aggregate) the entire economy. This includes assessment of difficult-to-measure Scope 3 emissions in a product's upstream supply chain, as illustrated in Figure 2. Furthermore, the EEIO method minimizes truncation errors that are common in process-based life cycle assessment (LCA).

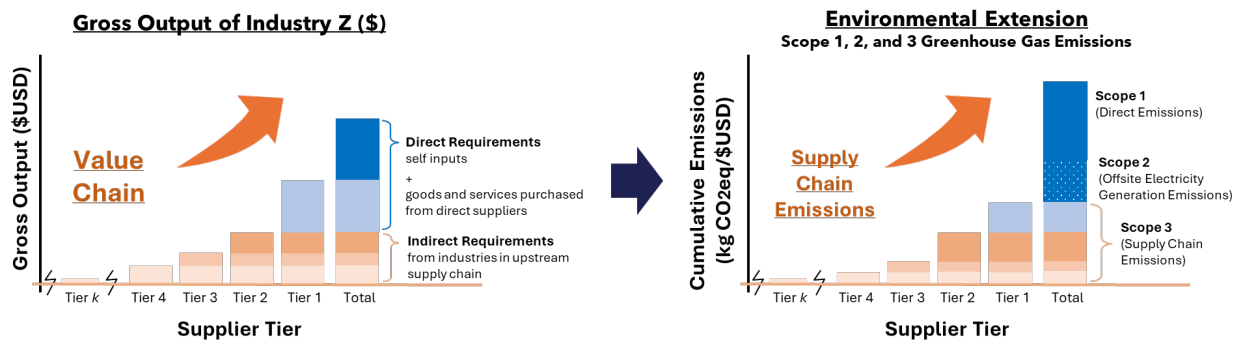

*Figure 2. Environmental extension of an economic IO model to quantify greenhouse gas emissions in product supply chains*

Despite the advantages of the economic foundation of the EEIO method, the monetary basis is also one of its main limitations. EEIO methods generally ignore the physical quantities of goods produced: all environmental impacts are expressed with respect to an output of goods or services in monetary units (e.g., $USD). Physical quantities do not typically factor into an EEIO assessment. As a result, EEIO models characterize environmental impact intensities in monetary terms, such as kg $CO_2$e/$USD. This is true of all of the major EEIO models available for the United States, including the U.S. Environmental Protection Agency's (EPA's) USEEIO model,[6,7] Carnegie Mellon's EIO-LCA model,[8] and the newer U.S. Department of Energy's (DOE's) EEIO-IDA model and scenario analysis tool.[4,9] The authors are not aware of any U.S.-focused model that allows for calculation of environmental impact intensities in mass-based units such as kg $CO_2$e/kg product.

Adding a physical extension to an EEIO model would enable estimation of industry-level environmental impacts on a mass basis (e.g., kg $CO_2$-eq/kg) rather than only on a monetary basis, which would be useful for more accurate Scope 3 product life cycle assessments, decarbonization scenario modeling at the product and industry level, and other purposes. Unfortunately, a barrier to such an extension is the limited availability of U.S. goods production data in physical units. This research develops a rational framework for estimating physical production when primary data are sparse. We use the framework to construct a new dataset for physical goods production in the United States, consisting of production data for 233 individual industries across 25 industrial subsectors. Data quality, limitations, and planned integration with DOE's EEIO-IDA model are discussed.

## Review of Related Work

While EEIO models don't typically incorporate physical units, a related family of models known as physical input-output tables (PIOTs) use physical units in place of monetary units in IO tables. PIOTs express goods exchanges in terms of physical units, such as kilograms, in an input-output framework. Use of these tables could address some of the challenges associated with the current practice of using monetary units as a proxy for physical units to represent flows of physical materials—but unfortunately, PIOTs have not yet been widely adopted.[10] Presently, the availability and coverage of PIOT models for the United States is limited and incomplete. Furthermore, PIOTs are rarely combined with EEIO models. Perhaps the most prominent example of a combined PIOT/EEIO model is EXIOBASE, a global multi-regional EEIO model that is available in both monetary and hybrid monetary/physical units.[11,12]

EXIOBASE datasets have been found to be in reasonable agreement with national statistics, LCA datasets, and other IO models in multiple studies,[13,14] though (unsurprisingly for such a large model), authors have also found that the level of agreement is imperfect.[15,16,17] EXIOBASE has a resolution (number of industries) higher than that of most national statistics used to build the model, requiring disaggregation assumptions that depend on analyst judgment. Results would therefore reasonably be expected to vary across models, and such variations should be properly attributed to uncertainty in both models rather than inaccuracy in either one. More importantly, data in EXIOBASE and similar models can only be as good as the national datasets from which the model is built. For mass-based physical goods production in the United States, national data are extremely limited. This study aims to fill that gap.

## Approach

### Physical Extension of the IO Model

Physical extension of an economic IO model typically involves constructing a physical input-output table (PIOT), which expresses goods exchanges in terms of physical units rather than monetary units in an input-output framework. The generation of a PIOT is a time- and data-intensive process requiring the use of detailed trade data to construct an entire input-output matrix representing the material exchanges. At the level of the U.S. economy, this would always result in a "hybrid" physical/monetary model because the economy produces a combination of goods and services. The output of an industry can only be expressed in physical units for the goods-producing industries (not the services). Therefore, the unit of output for the service-producing industries remains monetary, while the goods-producing industries are represented with a mass output unit. The multi-unit, hybrid nature of such a table complicates its use and makes this method difficult and inaccessible to nonexperts.

Here we suggest a simple approach that avoids the necessity of a hybrid IO table altogether in the physical IO extension. Rather than constructing a PIOT (a complete IO table that incorporates physical units), we instead physically extend the economic IO table in a manner analogous to EEIO environmental extension. To environmentally extend an IO table in the normal way, we leverage an environmental impact vector as shown in equation (1):

$$E = (I - A)^{-1} \cdot d \cdot r, \qquad (1)$$

where $E$ is the environmental impact vector (representative unit: kg $CO_2$e); $A$ is the Direct Requirements matrix (in monetary units) and $(I - A)^{-1}$ is the Leontief inverse of $A$; $d$ is the final demand vector (in monetary units); and $r$ is the environmental impact intensity (representative unit: kg $CO_2$e/$). Here, we say "representative unit" because this framework can be used to examine any environmental impact, not just greenhouse gas emissions. When many impacts are assessed, all of the environmental impact vectors are combined in the environmental satellite matrix, but each is still analyzed according to equation (1).

Analogously, to physically extend the IO table, we can use the same approach, using a physical production vector $P$ (in mass units such as kg) to compute a characteristic industry output price $p$ (in units of $/kg):

$$P = (I - A)^{-1} \cdot d \cdot (1/p). \qquad (2)$$

Combining equations (1) and (2), we can physically and environmentally extend the economic input output model, giving us the convenient unit we seek for environmental impact intensity (kg $CO_2$e/kg):

$$E = (I - A)^{-1} \cdot d \cdot (1/p) \cdot r^*, \qquad (3)$$

where $r^*$ is a modified impact intensity vector with a new unit of kg $CO_2$e/kg for all goods-producing industries in the model. The original units of kg $CO_2$e/$ will be maintained in $r^*$ for the service-producing industries. In this simple approach, the minimum data requirement to physically extend an EEIO model is a production vector $P$ of length $n$, containing the physical production values (in kg) of all goods-producing industries included in the model. For service-producing industries, monetary units can be maintained in $P$; in this case $p$ is unity and the manipulation has no effect.

### Quantifying Production in U.S. Industries

The US Bureau of Economic Analysis (BEA) releases the U.S. IO tables at the sector level (12 sectors) annually, summary level (71 sectors) annually, and detail level (over 400 sectors) every five years. In this study, we seek to quantify physical flows for the years 2018 and 2022 at both the summary level and the detail level. These years correspond to the survey years for the most recent Manufacturing Energy Consumption Survey (MECS) data release (2018) and the forthcoming MECS data release expected later this year (2022). These new datasets are intended for future integration with DOE's EEIO-IDA tool (Gause 2023) and alignment with ongoing decarbonization modeling efforts at DOE, some of which were recently published in DOE's *Industrial Decarbonization Roadmap*[18] and *Transformative Pathways for U.S. Industry: Unlocking American Innovation*[19] reports.

To satisfy the data requirements in the physical extension approach described in the previous section, primary-source production data from governmental or other high-quality sources are preferred. Unfortunately, production data in physical units from high-quality sources are extremely limited for the U.S. economy. The heatmap of Figure 3 illustrates our findings: of the 263 detail-level industries in goods-producing subsectors, we were able to locate high-quality production data with good industry coverage for only 48 (fewer than 20% of all goods-producing industries). For six additional industries, we were able to find data for selected products providing partial data coverage of the industry. Two of the industries within goods-producing subsectors provided support services and had no physical flows. That left 207 industries out of 263 (78%) with either no data found at all or very poor data coverage.

Considering the poor overall data coverage and the need for a complete dataset for physically extended EEIO modeling, alternative approaches were developed and used to estimate physical flows for cases in which primary-source data were unavailable. Physical flows were assessed using the following three methods, with the choice of principal data collection method selected depending on data availability and quality for an individual industrial subsector or goods-producing industry:

1. **Data-driven approach** (first choice, where data are available)

This method involves direct collection of physical production data from high-quality government sources such as the U.S. Geological Survey (USGS) or the Energy Information Administration (EIA), which can be used to provide estimates of mass directly. In some cases, detailed data were available for selected industries or products falling within a given subsector, but not for the full subsector. In such cases, partial data from primary sources can be combined with a complete sectoral dataset compiled using the price-driven approach (see below) to improve aggregated estimates and assess the data quality of the price-driven data for the industries with redundancy.

2. **Price-driven approach** (second choice, if data-driven method is not possible)

For industrial subsectors that do not have comprehensive physical output data available from government sources, mass is imputed by dividing the subsector's production in monetary units (available in the BEA tables[1]) by an assumed price for that subsector's output ($/kg). Prices are estimated using trade data from UN Comtrade, which reports cost, insurance, and freight (CIF) trade values (USD) and net weight (kg) by HS commodity codes,[2] and USA Trade, which provides calculated duties by HS commodity codes (the latter being necessary to convert the CIF values extracted from Comtrade to producer prices, aligning with IO table values).

Comtrade includes price information on both imports and exports, with data reported by HS commodity code. For the United States, import data are preferred to export data (when using as an estimate of domestic prices) due to the nature of the U.S. economy as a net importer. HS commodity codes for imports are translated to NAICS codes[3] and then to BEA industry groupings[4] in Python using crosswalk concordance files available from the U.S. Census Bureau.[5] The CIF values, net weight, and calculated duty data are then aggregated according to BEA industry groupings to generate a characteristic price for each subsector. To convert the CIF values to producer price values, the following steps were completed:

   a. For each industry, the calculated duty value (from USA Trade) was divided by the CIF values (from UN Comtrade) to output a duty rate.
   b. Three duty rate tiers were established by identifying the minimum and maximum duty rates across all industries with a non-zero duty rate and defining three equally sized intervals (low, medium, and high duty) to span the range between min/max.
   c. Industries with duty rates falling within the limits of the upper interval were defined as "high duty," industries with duty rates falling within the central interval were defined as "medium duty," and industries with duty rates within the lower interval were defined as "low duty." Industries were not necessarily evenly distributed across the three intervals.

---

[1] BEA tables being used for production detail (in monetary units) include "The Use of Commodities by Industries, After Redefinitions (Producer's Prices) - Summary" and "Gross Output by Industry" for the year 2018 (BEA 2024a; 2024b).
[2] Harmonized System (HS) Codes: an industry classification system administered by the World Customs Organization (WCO) that classifies traded products. Commodities are assigned six-digit codes.
[3] North American Industry Classification System (NAICS): an industry classification developed under the Office of Management and Budget (OMB) that classifies sectors, subsectors, industries, and commodities of the U.S. business economy. These are organized by 2-, 3-, 4-, and 6-digit codes, respectively.
[4] Bureau of Economic Analysis (BEA) industry groupings: classification system used by BEA to analyze industry data, including the input-output accounts. These codes are based on NAICS with some minor differences in aggregation of subsectors.
[5] U.S. Census Bureau: The U.S. Census Bureau provides crosswalk concordance files to translate HS codes to NAICS codes.

d. An average duty value was computed for each duty tier, calculated as the mean duty rate for all industries grouped within each duty tier.
e. Industries within each duty tier were assigned the average duty rate for that tier. For the sectors with a duty rate value of zero in USA Trade, no duty was assigned.
f. The basic price for each industry was calculated from UN Comtrade CIF and net weight data, combined with the assigned duty rate, using equation (4):

$$Basic\ Price = \frac{CIF + CIF * Assigned\ Duty\ Rate}{Net\ Weight\ (kg)}. \tag{4}$$

g. Finally, the basic price is converted to producer prices using a ratio of basic to producer prices extracted from the Tau matrix[20] (i.e., the basic-to-producer-price ratio matrix) in USEEIO v2.3:

$$Producer\ Price = \frac{Basic\ Price}{Tau\ ratio}. \tag{5}$$

Physical values are then calculated by dividing the subsector-specific Gross Output in USD by the subsector producer price.

When partial data on physical flows are available from government sources (i.e., the data-driven method), these data are compared and used to assess data quality for the price-driven estimates (see Data Collection Method Comparison). In future iterations, redundant datasets may be further integrated to compose a "best available" dataset to improve data quality.

3. **Input-driven approach**:

The final method for assessing physical production, input-driven, is considered the least accurate but is used when neither of the previous two methods are possible, which is true (e.g.) for the Construction subsector. We are forced to use the input-driven method if a subsector produces or transforms physical goods, but no high-quality sources are available for physical production (precluding the use of the data-driven method) and the subsector does not participate in international trade (precluding the use of the price-driven method). Prices in these industries are estimated based on a mass balance as the sum of inputs, using equation (6), which is modified from an expression in the UN SUT Handbook[21] to incorporate a waste coefficient:

$$p_i = \frac{x_i}{(1-w_i)\sum_{j=1}^{n} z_{i,j}}, \tag{6}$$

where $p$ is the characteristic price of industry $i$, $x_i$ is industry $i$'s total output in dollars, $z_{i,j}$ is the mass of the inputs from sector $j$ used by sector $i$, and $w_i$ is a waste coefficient for industry $i$. The waste coefficient is a value between zero and one that accounts for material that is provided as an input but leaves the mass-balance as waste without being incorporated into the industry's final products (e.g., construction scrap sent to landfill).

| Subsector | | | | | | | | | | | | |
|---|---|---|---|---|---|---|---|---|---|---|---|---|
| **111CA Farms** | Oilseed farming | Grain farming | Vegetable and melon farming | Fruit and tree nut farming | Greenhouse, nursery, and floriculture production | Other crop farming | Beef cattle ranching and farming | Dairy cattle and milk production | Other animal production | Poultry and egg production | | |
| **113FF Forestry & Fishing** | Forestry and logging | Fishing, hunting and trapping | Support activities for agriculture and forestry (N/A, no physical flows) | | | | | | | | | |
| **211 Oil & Gas** | Oil and gas extraction | | | | | | | | | | | |
| **212 Mining** | Coal mining | Iron, gold, silver, and other metal ore mining | Copper, nickel, lead, and zinc mining | Stone mining and quarrying | Other nonmetallic mineral mining and quarrying | | | | | | | |
| **23 Construction** | Health care structures | Educational and vocational structures | Office and commercial structures | Power and communication structures | Transportation and highways and streets | Manufacturing structures | Other nonresidential structures | Single-family residential | Multifamily residential | Other residential | Nonresidential repair | Residential repair |
| **321 Wood Products** | Sawmills and wood preservation | Veneer, plywood, and engineered wood | Millwork | All other wood products | | | | | | | | |
| **327 Nonmet. Minerals** | Clay product and refractory | Glass and glass products | Cement | Ready-mix concrete | Concrete pipe, brick, and block | Other concrete products | Lime and gypsum products | Abrasive products | Cut stone and stone products | Ground or treated mineral and earth | Mineral wool | Misc. nonmetallic mineral products |
| **331 Primary Metals** | Iron and steel mills and ferroalloys | Steel products from purchased steel | Alumina refining and primary aluminum production | Secondary smelting and alloying of aluminum | Aluminum products from purchased aluminum | Nonferrous metal (except Al) smelting and refining | Copper rolling, drawing, extruding and alloying | Nonferrous metal (except Cu, Al) rolling, drawing, extruding | Ferrous metal foundries | Nonferrous metal foundries | | |
| **332 Fabricated Metals** | All other forging, stamping, and sintering | Custom roll forming | Metal crown, closure, and other metal stamping (except automotive) | Cutlery and handtool manufacturing | Plate work and fabricated structural product manufacturing | Ornamental and architectural metal products manufacturing | Power boiler and heat exchanger manufacturing | Metal tank (heavy gauge) manufacturing | Metal can, box, and other metal container (light gauge) manufacturing | Hardware manufacturing | Spring and wire product manufacturing | Machine shops |
| | Turned product and screw, nut, and bolt manufacturing | Coating, engraving, heat treating and allied activities | Valve and fittings other than plumbing | Plumbing fixture fitting and trim manufacturing | Ball and roller bearing manufacturing | Ammunition, arms, ordnance, and accessories manufacturing | Fabricated pipe and pipe fitting manufacturing | Other fabricated metal manufacturing | | | | |
| **333 Machinery** | Farm machinery and equipment | Lawn and garden equipment | Construction machinery | Mining and oil and gas field machinery | Semiconductor machinery | Other industrial machinery | Optical instruments and lenses | Photographic and photocopying equipment | Other commercial and service industry machinery | Industrial and commercial fan and blower and air purification equipment | Heating equipment (except warm air furnaces) | Air conditioning, refrigeration, and warm air heating equipment |
| | | Industrial molds | Machine tools | Special tool, die, jig, and fixture manufacturing | Cutting and machine tool accessory, rolling mill, and other metalworking machinery | Turbine and turbine generator set units | Speed changers, industrial high-speed drives, and gears | Mechanical power transmission equipment | Other engine equipment manufacturing | Measuring, dispensing, and other pumping equipment manufacturing | Air and gas compressor manufacturing | Material handling equipment manufacturing | Power-driven handtool manufacturing |
| | Other general purpose machinery manufacturing | Packaging machinery manufacturing | Industrial process furnace and oven manufacturing | Fluid power process machinery | | | | | | | | |
| **334 Comp. & Electronics** | Electronic computer manufacturing | Computer storage device manufacturing | Computer terminals and other computer peripheral equipment | Telephone apparatus manufacturing | Broadcast and wireless communications equipment | Other communications equipment manufacturing | Semiconductor and related device manufacturing | Printed circuit assembly (electronic) manufacturing | Other electronic component manufacturing | Electromedical and electrotherapeutic apparatus manufacturing | Search, detection, and navigation instruments manufacturing | Automatic environmental control manufacturing |
| | Industrial process variable instruments manufacturing | Totalizing fluid meter and counting device manufacturing | Electricity and signal testing instruments manufacturing | Analytical laboratory instrument manufacturing | Irradiation apparatus manufacturing | Watch, clock, and other measuring and controlling device manufacturing | Audio and video equipment manufacturing | Manufacturing and reproducing magnetic and optical media | | | | |
| **335 Electr. & Appliances** | Electric lamp bulb and part manufacturing | Lighting fixture manufacturing | Small electrical appliance manufacturing | Major household appliance manufacturing | Power, distribution, and specialty transformer manufacturing | Motor and generator manufacturing | Switchgear and switchboard apparatus manufacturing | Relay and industrial control manufacturing | Storage battery manufacturing | Primary battery manufacturing | Communication and energy wire and cable manufacturing | Wiring device manufacturing |
| | Carbon and graphite product manufacturing | All other miscellaneous electrical equipment and components | | | | | | | | | | |
| **3361MV Motor Vehicles & Parts** | Automobile manufacturing | Light truck and utility vehicles | Heavy duty trucks | Motor vehicle body | Truck trailers | Motor homes | Travel trailer and campers | Motor vehicle gasoline engine and engine parts | Motor vehicle electrical and electronic equipment | Motor vehicle steering, suspension (except spring), and brake systems | Motor vehicle transmission and power train parts | Motor vehicle seating and interior trim |
| | Motor vehicle metal stamping | Other motor vehicle parts | | | | | | | | | | |
| **3364OT Other Transport. Equip.** | Aircraft manufacturing | Aircraft engine and engine parts manufacturing | Other aircraft parts and auxiliary equipment manufacturing | Guided missile and space vehicle manufacturing | Propulsion units and parts for space vehicles and guided missiles | Railroad rolling stock manufacturing | Ship building and repairing | Boat building | Motorcycle, bicycle, and parts manufacturing | Military armored vehicle, tank, and tank component manufacturing | All other transportation equipment manufacturing | |
| **337 Furniture** | Wood kitchen cabinet and countertops | Upholstered household furniture | Nonupholstered wood household furniture | Other household nonupholstered furniture | Institutional furniture | Office furniture and custom architectural woodwork and millwork | Showcase, partition, shelving, and lockers | Other furniture related products | | | | |
| **339 Misc. Manufact.** | Surgical and medical instruments | Surgical appliance and supplies | Dental equipment and supplies | Ophthalmic goods | Dental laboratories | Jewelry and silverware | Sporting and athletic goods | Doll, toy, and games | Office supplies (except paper) | Signs | All other miscellaneous products | |
| **311FT Food & Beverage** | Dog and cat food | Other animal food | Flour milling and malt | Wet corn milling | Soybean and other oilseeds | Fats and oils | Breakfast cereal | Sugar and confectionery products | Frozen foods | Fruit and vegetable canning, pickling, and drying | Fluid milk and butter | Cheese |
| | Dry, condensed, and evaporated dairy products | Ice cream and frozen desserts | Animal (except poultry) slaughtering, rendering, and processing | Poultry processing | Seafood product preparation and packaging | Bread and bakery products | Cookie, cracker, pasta, and tortilla | Snack food | Coffee and tea | Flavoring syrup and concentrate | Seasoning and dressing | All other foods |
| | Soft drink and ice | Breweries | Wineries | Distilleries | Tobacco | | | | | | | |
| **313TT Textile Mills** | Fiber, yarn, and thread | Fabric | Textile and fabric finishing and fabric coating | Carpet and rug | Curtain and linen | Other textile product | | | | | | |
| **315AL Apparel & Leather** | Apparel | Leather and allied products | | | | | | | | | | |
| **322 Paper** | Pulp mills | Paper mills | Paperboard mills | Paperboard containers | Paper bag and coated and treated paper | Stationery products | Sanitary paper products | All other converted paper products | | | | |
| **323 Printing** | Printing | Support activities for printing (N/A, no physical flows) | | | | | | | | | | |
| **324 Petroleum & Coal** | Petroleum refineries | Asphalt paving mixture and blocks | Asphalt shingle and coating materials | Other petroleum and coal products | | | | | | | | |
| **325 Chemicals** | Petrochemicals | Industrial gases | Synthetic dye and pigments | Other basic inorganic chemicals | Other basic organic chemicals | Plastics material and resins | Synthetic rubber and artificial and synthetic fibers and filaments | Medicinals and botanicals | Pharmaceutical preparation | In-vitro diagnostic substances | Biological products (except diagnostic) | Fertilizers |
| | Pesticide and other agricultural chemicals | Paint and coatings | Adhesives | Soap and cleaning compounds | Toilet preparations | Printing ink | All other chemical product and preparations | | | | | |
| **326 Plastics & Rubber Prod.** | Plastics packaging materials and unlaminated film and sheets | Plastics pipe, pipe fitting, and unlaminated profile shapes | Laminated plastics plate, sheet (except packaging), and shapes | Polystyrene foam products | Urethane and other foam products (except polystyrene) | Plastics bottles | Other plastics products | Tire | Rubber and plastics hoses and belting | Other rubber products | | |

*Figure 3. Availability of U.S. physical production data (in mass units such as kg) from high-quality primary sources, by subsector*

## Discussion

Data collection and analysis following the framework outlined in this paper is now in progress to compile a preliminary physical flows dataset for the United States. An important consideration for this work is data quality. One way we can assess the overall quality of our framework is to compare results across more than one of the proposed estimation methods (i.e., data-driven, price-driven, and input-driven) in cases where more than one data collection method is possible. Such data redundancy is present for the industries shaded green ("good data coverage") in Figure 3 because in these industries, sufficient data exists for both the data-driven and the price-driven methods to be used.

Two such subsectors are Crop Production (111CA) and Petroleum and Coal Products (324). The mass values (million metric tons) for these two subsectors are compared in Figure 4. The percent difference between mass values estimated with the two methods ranged from 17-167%. While this is a greater difference than we would prefer, the order of magnitude matches for the two completely independent methods, which represents significant progress towards a usable dataset. Work will continue to evaluate possible sources of error and opportunities for method improvement.

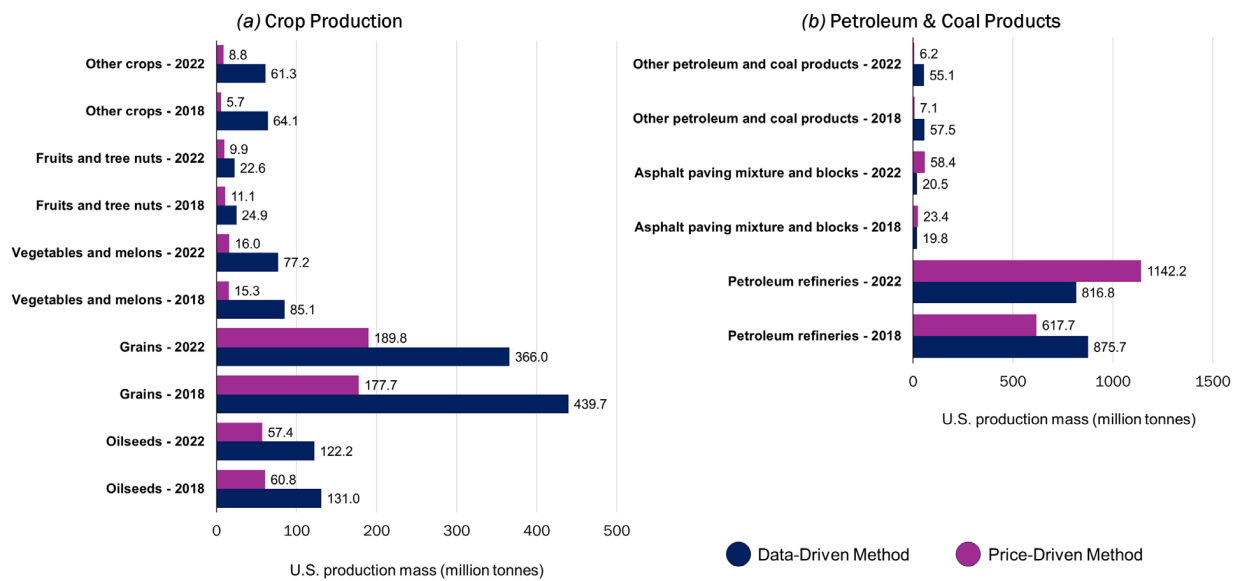

*Figure 4. Comparison of physical production estimates for the data-driven and price-driven methods for two industrial subsectors: Crop Production and Petroleum and Coal Products.*

A key issue for this project is that the data-driven approach is preferred (based on data quality) but data coverage is normally insufficient to apply this method to an entire subsector. To supplement this gap, the trade-based imputation (price-driven) approach is used. However, differences in estimates made using these two methods can be substantial and vary depending on the subsector. Further, different factors drive the potential inaccuracies of each of these methods. For example, inaccuracy in governmental production data (data-driven method) could be driven by incomplete reporting, whereas inaccuracy in production estimates made using the price-driven method could be driven by a wide range of factors such as price volatility, price heterogeneity between goods produced for domestic consumption vs. import or export, tariffs and other trade policies.

## Conclusions & Recommendations

We have offered here a simple framework for the simultaneous physical and environmental extension of an economic input-output model that minimizes data requirements and simplifies model structure compared to a PIOT-based approach. The developed framework is now being used to develop a physical flows dataset for integration with the U.S. Department of Energy's (DOE's) EEIO for Industrial Decarbonization (EEIO-IDA) scenario modeling tool. Physical production data is needed for every goods-producing industry in the U.S. economy to enable the physical IO extension. This has proved extremely challenging due to the scarcity of high-quality raw data and poor harmonization across industry classification systems used in data reporting. Few industries have complete data coverage from primary governmental sources, meaning that an alternative price-driven (price imputation) estimation method is often the best available approach for assessing physical flows. Future work is focused on building a complete physical flows dataset for the United States for the years 2018 and 2022 and performing a comprehensive data quality assessment, highlighting areas of challenge and opportunity for mass-based EEIO.

## Acknowledgments

This work is based upon funding from the Alliance for Sustainable Energy, LLC, Managing and Operating Contractor for the National Renewable Energy Laboratory for the U.S. Department of Energy.

## About the Authors

**Heather Liddell** is an Assistant Professor at Purdue University with a joint appointment in Mechanical Engineering and Environmental & Ecological Engineering. Her research focuses on technology and systems strategies to improve resource utilization and reduce adverse environmental impacts of manufacturing. Liddell holds BS and MS degrees in mechanical engineering and a PhD in materials science from the University of Rochester.

**Beth Kelley** is a Graduate Research Assistant at Purdue University. Her research focuses on quantifying mass flows in the U.S. manufacturing sector and developing a database of mass-based emission factors (kg $CO_2$e/kg) for U.S. goods production. She completed her BS in May 2023 and expects to graduate with her MS in May 2025 from the Environmental & Ecological Engineering program at Purdue University.

**Liz Wachs** is a Researcher at the National Renewable Energy Laboratory, focusing on industrial decision making for sustainability. Liz holds a BA in ancient history and classical civilization from the University of Texas at Austin, an MA in Desert Studies from Ben Gurion University of the Negev, a BS in chemical engineering and PhD in agricultural and biological engineering from Purdue University.

**Alberta Carpenter** is a Distinguished Member of Research Staff and a Senior Researcher at the National Renewable Energy Laboratory and leads NREL's strategic analysis work focusing on sustainable manufacturing to include industrial decarbonization, circular economy, environmental and justice impacts.

**Joe Cresko** is the Chief Engineer for DOE's Industrial Efficiency and Decarbonization Office, where he leads the Strategic Analysis Team. This team helps to assess the life cycle and cross-sectoral impacts of new manufacturing advances. Cresko holds a BS in chemical engineering from Bucknell University and an MS in engineering science and mechanics from Penn State University.